# Assessing the role of ITER ECE's oblique view in resolving non-thermal emissions

Saeid Houshmandyar and William L. Rowan

*Institute for Fusion Studies, The University of Texas at Austin, Austin, TX 78712, USA*

## Abstract

Systematic discrepancies between electron temperature ($T_e$) measurements from radially viewing electron cyclotron emission (ECE) and Thomson scattering (TS) diagnostics have been observed in multiple tokamaks and are widely attributed to non-Maxwellian features in the electron velocity distribution function (EVDF). As the International Thermonuclear Experimental Reactor (ITER) is expected to operate at much higher core temperatures than present devices, accurate $T_e$ measurements from ECE become increasingly critical, particularly in the presence of non-Maxwellian EVDFs. This work presents ECE spectra simulations performed to assess the diagnostic capability of the ITER oblique view under ITER conditions where non-Maxwellian distributions are present. The results show that above a critical oblique angle, Doppler broadening becomes the dominant effect, masking fine-scale spectral signatures of non-thermal distortions across a wide range of conditions. Furthermore, at higher EC harmonics in either polarization, the spectra remain largely unaffected by non-thermal emissions, enabling reliable reconstruction of $T_e$-profiles. These findings demonstrate that the ITER oblique ECE view retains sufficient sensitivity to non-thermal electrons while providing robust and accurate $T_e$-profile measurements under reactor-relevant conditions. The results presented here also have direct implications for ITER ECE system operation and channel selection, particularly in the presence of non-Maxwellian electron populations.

## 1. Introduction

Electron cyclotron emission (ECE) diagnostics are widely used in contemporary tokamaks and stellarators, to provide high temporal and spatial resolution measurements of electron dynamics. ECE systems are not only used for determining electron temperature ($T_e$)[1,2] and its fluctuations[3] but also play a key role in diagnosing magnetohydrodynamic (MHD) instabilities,[4] magnetic island detection[5,6] and enabling plasma control applications.[7,8] Although Thomson Scattering (TS) diagnostics also provide detailed $T_e$-profiles, the importance of ECE becomes increasingly evident where in-vessel access for diagnostics such as TS may be severely limited and where long-term operation of optical components poses additional challenges in reactor-grade tokamaks.

Despite the widespread use of ECE diagnostics,[9,10,11,12,13,14,15,16] a key challenge remains: understanding how to accurately and reliably measure $T_e$ using ECE systems. This need arises from longstanding, systematic, and well-documented discrepancies between temperatures measured by ECE ($T_{ECE}$) and those measured by Thomson scattering ($T_{TS}$). These discrepancies become particularly pronounced at high bulk plasma temperatures. Early experiments at the Tokamak Fusion Test Reactor (TFTR)[17] and the Joint European Torus (JET)[18] established a threshold of approximately 7 keV, beyond which the ECE and TS temperature measurements diverged. In these early observations, ECE measurements typically yielded higher electron temperatures than TS.

However, subsequent observations introduced more complex behavior. For example, in exclusively ICRF-heated plasmas at the Alcator C-Mod tokamak, good agreement was observed between ECE and TS measurements,[19] though it is suspected that the threshold at which discrepancies emerge was not reached. In contrast, at the Frascati Tokamak Upgrade (FTU), where electron cyclotron heating (ECH) was used, $T_{TS}$ were reported to exceed those from ECE.[20] More recently, the JET deuterium-tritium (DT) campaigns, employing both neutral beam injection (NBI) and ICRF heating, revealed that $T_{ECE}$ can be either higher or lower than $T_{TS}$ depending on the discharge. These experiments also reported discrepancies between the second and third harmonic ECE spectra in the extraordinary mode.[21]

Such findings have led to the interpretation that non-Maxwellian features[22] in the electron velocity distribution function (EVDF) can significantly affect the electron temperature inferred from ECE measurements. These features may arise directly from specific heating mechanisms or indirectly through modifications of the background plasma, for example due to fast-ion populations that distort the electron distribution. The impact of these effects can vary across different ECE harmonics and viewing geometries. These effects may become even more significant in the International Thermonuclear Experimental Reactor (ITER), where the central electron temperature is projected to reach approximately 25 keV.

ITER's diagnostic mission aims to support reliable plasma operation, ensure machine protection, and advance the scientific understanding critical for future fusion power plants (FPPs). The ECE diagnostic contributes to this mission by providing non-intrusive measurements and enabling cross-validation with other diagnostics, such as TS. Therefore, resolving the ECE-TS temperature discrepancy is essential to ensure accurate $T_e$ measurements in ITER and beyond.

To address this issue and study the effect of non-Maxwellian distributions, a 9.25° oblique ECE view was incorporated into the quasi-optical (QO) design[23] of the ITER ECE diagnostics. Figure 1 shows the second diagnostic shielding module (DSM2) of equatorial port 9 (EP9) of ITER, where the ECE QO path is located. Previous studies[24] has demonstrated that an oblique ECE view will enhance sensitivity to non-thermal electrons. These simulations showed that while perpendicular lines of sight primarily probe electrons below ~10 keV, oblique views can extend sensitivity up to ~70 keV.

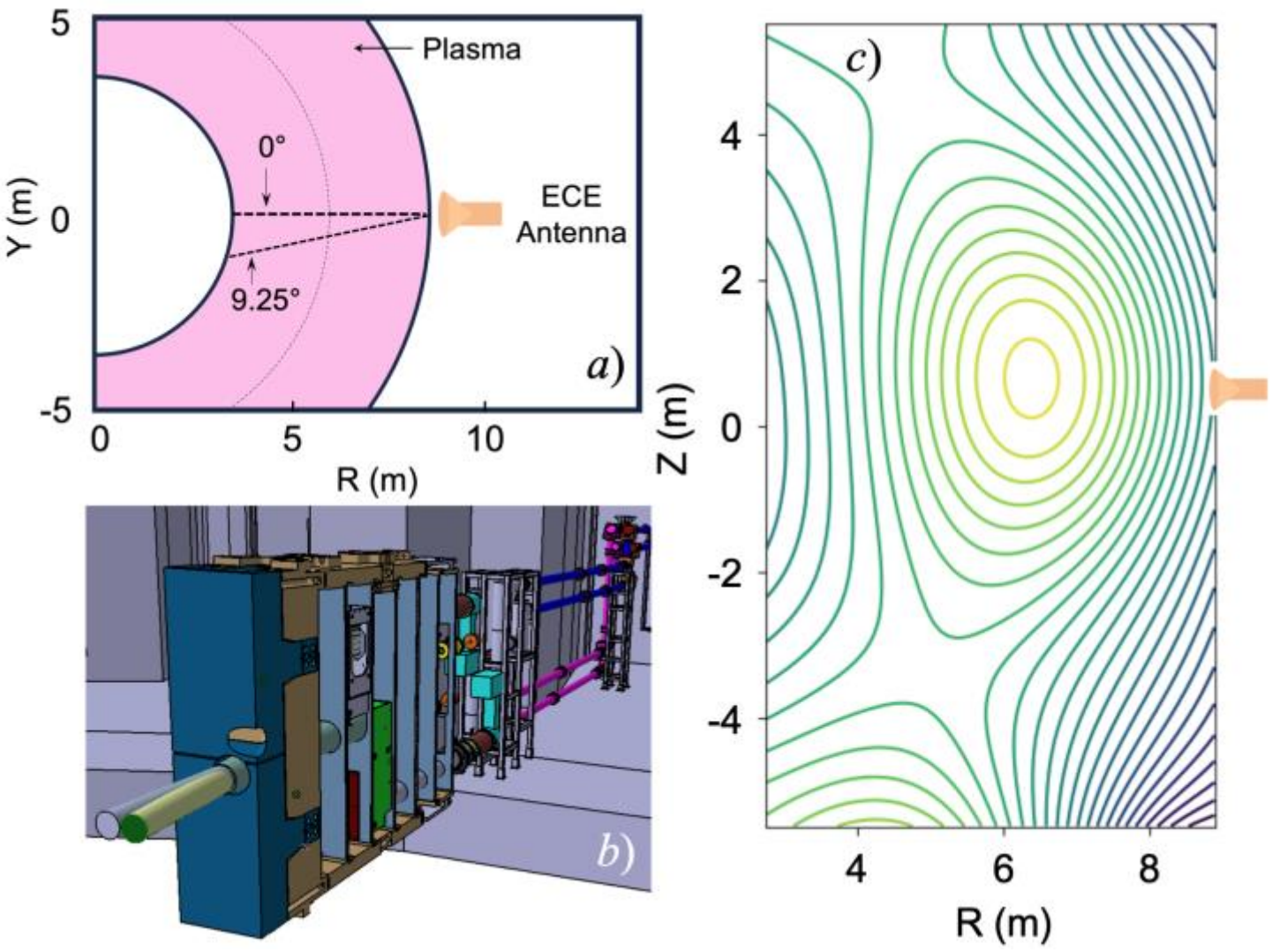

**Figure 1.** *a*) Top view of the ITER tokamak showing the relative locations of the ECE antennas for the radial (0°) and oblique (9.25°) lines of sight. *b*) Front view of the second Diagnostic Shielding Module (DSM2) in the equatorial port 9 (EP9), as seen from inside the ITER vacuum vessel, showing the quasi-optical design of the ITER ECE diagnostics. The calculated beam widths for both the radial and oblique views are shown as protruding tubes. Only the antenna for radial view is shown for clarity, but each view has its own dedicated antenna. *c*) Magnetic equilibrium of the ITER discharge 130506 at $t$ = 230 s, with the location of the ECE antennae with respect to the flux surfaces.

The present work utilizes multiple non-Maxwellian EVDF models to predict the resulting ECE spectra, with the goal of assessing the effectiveness of ITER's oblique ECE view when non-thermal electrons dilute the bulk distribution. Although such EVDF distortions can arise from various plasma heating mechanisms, the present study does not attempt to model those mechanisms or their influence on the distribution function. Instead, it focuses on determining whether the implemented oblique viewing angle in ITER is sufficient to address potential ECE-TS temperature discrepancies and whether it retains sensitivity to high-energy electrons. We also evaluate its fidelity in reconstructing Maxwellian distributions at elevated temperatures and examine how non-Maxwellian features in the EVDF may affect the interpretation of ECE measurements. Ultimately, the objective is to assess how reliably ECE diagnostics can report $T_e$-profiles in high-temperature, reactor-grade plasmas.

The novelty of this work lies in providing a systematic assessment of ITER's implemented 9.25° oblique ECE view using ITER IMAS (Integrated Modelling & Analysis Suite) profiles and controlled, parameterized non-Maxwellian EVDF distortions. In contrast to previous studies, which examined idealized viewing angles or EVDF modifications tied to specific heating scenarios, the present analysis isolates the diagnostic response itself by scanning both the radial extent and momentum-space location of non-Maxwellian features. This approach identifies the harmonics and polarization combinations that remain reliable for $T_e$-profile reconstruction under reactor-grade conditions and establishes quantitative expectations for when the oblique view suppresses non-thermal emissions. These results offer diagnostic guidance directly applicable to ITER operation and extend beyond earlier oblique-view analyses in tokamak plasmas.

## 2. Radiation transport modelling

To support the analysis presented in this work, simulations were carried out using kinetic profiles and magnetic equilibrium from a recent ITER IMAS H-mode scenario at full field (5 Tesla). The corresponding kinetic profiles are shown in Figure 2.*a*, and the magnetic equilibrium is shown in Figure 1.*c*. Figure 2.*b* includes the profiles of the first three electron cyclotron harmonics, as well as the right-hand, left-hand, upper-hybrid, and lower-hybrid cutoff and resonance frequencies, based on cold plasma resonances; relativistic effects are included in the emissivity and absorption calculations used in the ECE modelling, presented later in this section. The high frequency radiometer for ITER ECE is designed to measure the second-harmonic X-mode in the 220-340 GHz range. As depicted in Figure 2.*b*, this radiometer frequency range is free of cutoffs and avoids spectral overlap with the third-harmonic EC emission, ensuring clean measurement of the intended harmonic.

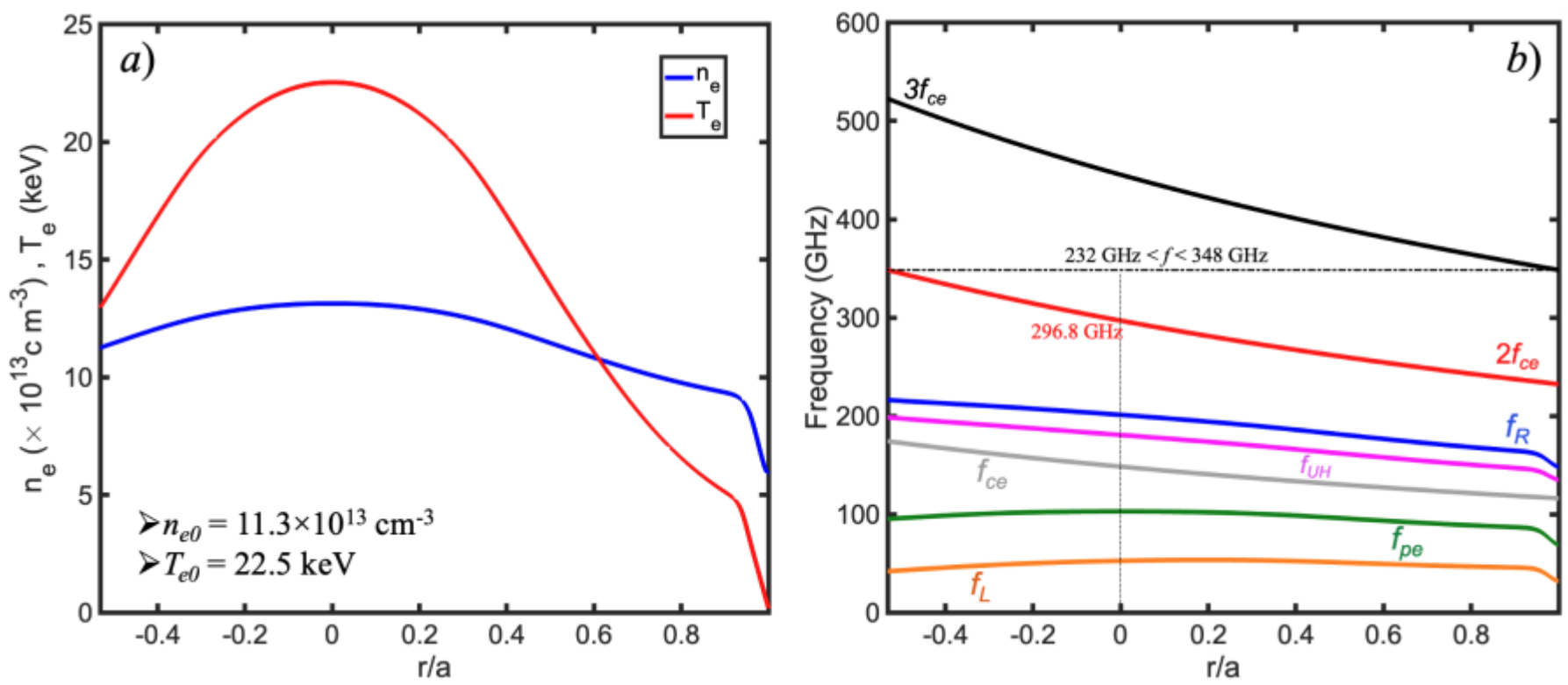

**Figure 2.** *a*) Kinetic profiles for ITER discharge 130506 (full field, $B_0 = 5.3$ Tesla) at $t = 203$ s. *b*) Profiles of the first three electron-cyclotron harmonics, as well as the relevant cutoff, resonance, plasma, and cyclotron frequencies, plotted as a function of normalized radius.

As described in Appendix I, for a given ECE frequency under the cold plasma approximation, the localization of channels is straightforward. Therefore, by accounting for the relativistic down-shift of the channels, the localization of electron kinetic energy on the radial profile can be defined as[22]

$$KE = m_e c^2(\gamma - 1) = m_e c^2\left(\frac{R_{cp}}{R_{rel}} - 1\right). \tag{1}$$

Here, $\gamma$ is the Lorentz factor, $c$ is the speed of light, $m_e$ is the electron mass, and $R_{rel}$ and $R_{cp}$ are ECE radial locations based with and without relativistic effects correction as described in Appendix I.

The emissivity function peaks at different normalized momenta, $u/u_{\text{th}}$ where $u = \gamma m v$ and $u_{\text{th}}$ is the electron thermal momentum.[21] Figure 3 illustrates how the peak and width of the emissivity function depend on electron temperature when expressed in normalized momentum space, $u/u_{\text{th}}$. As a result, emission at a given EC harmonic preferentially samples a finite region of the EVDF centered around the corresponding emissivity peak. For the temperature range relevant to the present ITER scenario, the second- and third-harmonic emission primarily probes electrons in the approximate range $0.5 < u/u_{th} < 3.2$. At the higher core temperatures expected in ITER, where ECE-TS discrepancies are most relevant, the emissivity layer becomes narrower in normalized momentum, reducing the sampled region to approximately $0.5 < u/u_{\mathrm{th}} < 1.5$.

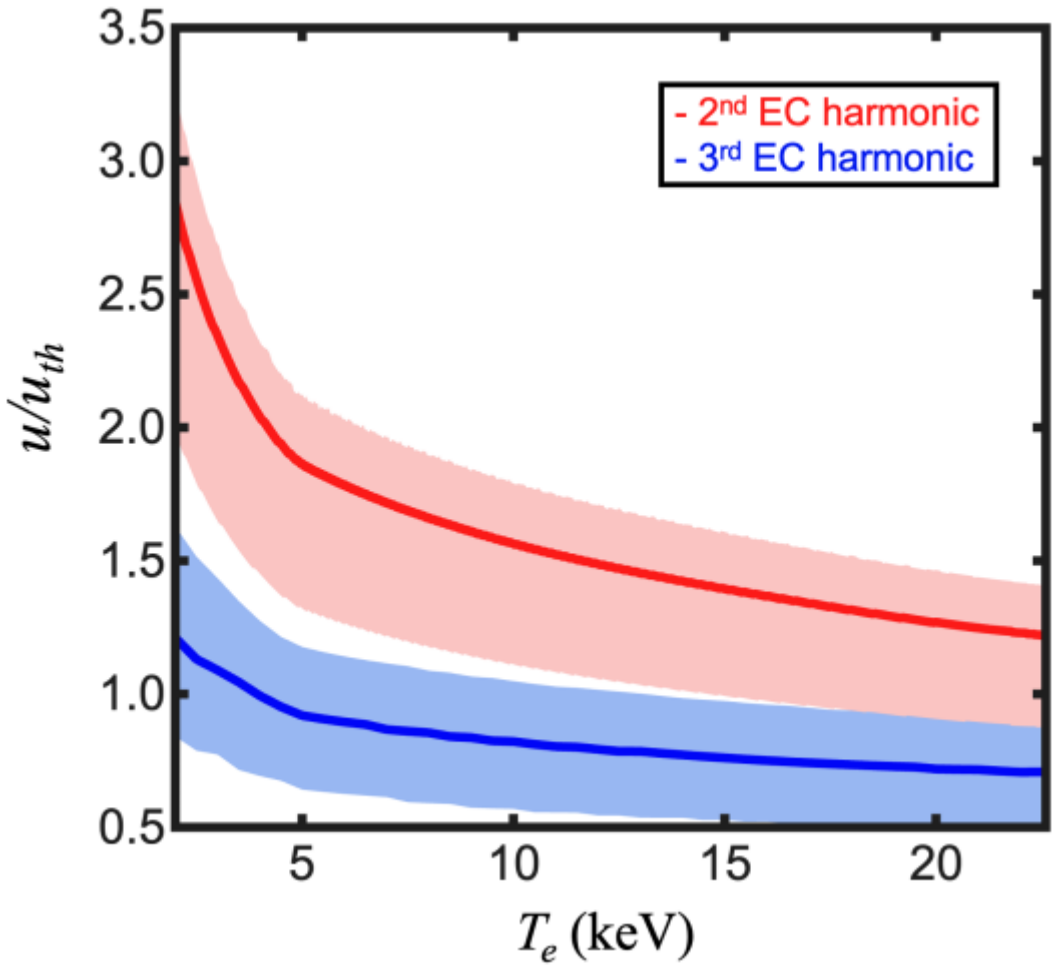

**Figure 3.** Momentum normalized to the thermal momentum at the peak of the emissivity function for the second (blue) and third (red) ECE harmonic domains. The error bars represent the width of the emissivity layer, defined by the positions $R_{\mathrm{rel}}$-R05 and $R_{\mathrm{rel}}$+R95, as they are detailed in Appendix I.

ECE Modelling in this work was performed using GENRAY, a three-dimensional ray-tracing code that simulates emission, absorption, and radiation transport for both thermal and nonthermal electron distributions.[25] At each point along rays originating from the plasma, GENRAY computes the emission and absorption coefficients and solves the radiation transport equation back to the antenna. Ray trajectories are calculated using the magnetized cold-plasma dispersion relation to determine the refractive index and propagation path. Relativistic effects are included in the calculation of emissivity, absorption, and line broadening through the relativistic dielectric tensor and radiation transport formulation. This separation allows accurate treatment of wave propagation while retaining fully relativistic emission physics.

To calculate the absorption (*i.e.*, the imaginary part of the perpendicular refractive index), the anti-Hermitian component of the relativistic dielectric tensor was computed, following Harvey *et al*[25]. As shown in Equations (6)-(8) of reference 25, the anti-Hermitian tensor not only enforces the resonance condition for ECE emission but also includes a term proportional to the derivative of the momentum distribution function with respect to perpendicular velocity ($v_\perp$). For a Maxwellian distribution function, this derivative is always negative, resulting in positive emissivity. As described in Appendix I, the radiation temperature ($T_{rad}$) behaves as $T_e \cdot e^{-\int \alpha\, ds}$. This implies that $T_{rad}$ decays toward the local electron temperature at the ray's origin. However, if the derivative becomes positive, $T_{rad}$ increases along the trajectory, leading to amplified radiation and the appearance of non-thermal emissions.

Here, the ECE modelling for Maxwellian distributions uses isotropic relativistic Maxwellian EVDF described by

$$f_m = \frac{1}{4\pi m_e^2 c T_e K_2(\Theta)} e^{-\Theta\gamma} \,. \qquad (2)$$

In Eq. 2, $\Theta = m_e c^2/T_e$, $T_e$ is obtained at each radius from the input profiles, and $K_2$ is the Macdonald function.[26] Figure 4 shows the measured emission flux spectra at the plasma edge for extraordinary and ordinary modes with a Maxwellian EVDF, for both radial and oblique ITER ECE views. The spectra are plotted versus emission frequency normalized to the on-axis fundamental cyclotron frequency ($f_{\mathrm{ec0}}$ = 148.4 GHz), per unit energy range of the emitting electrons.

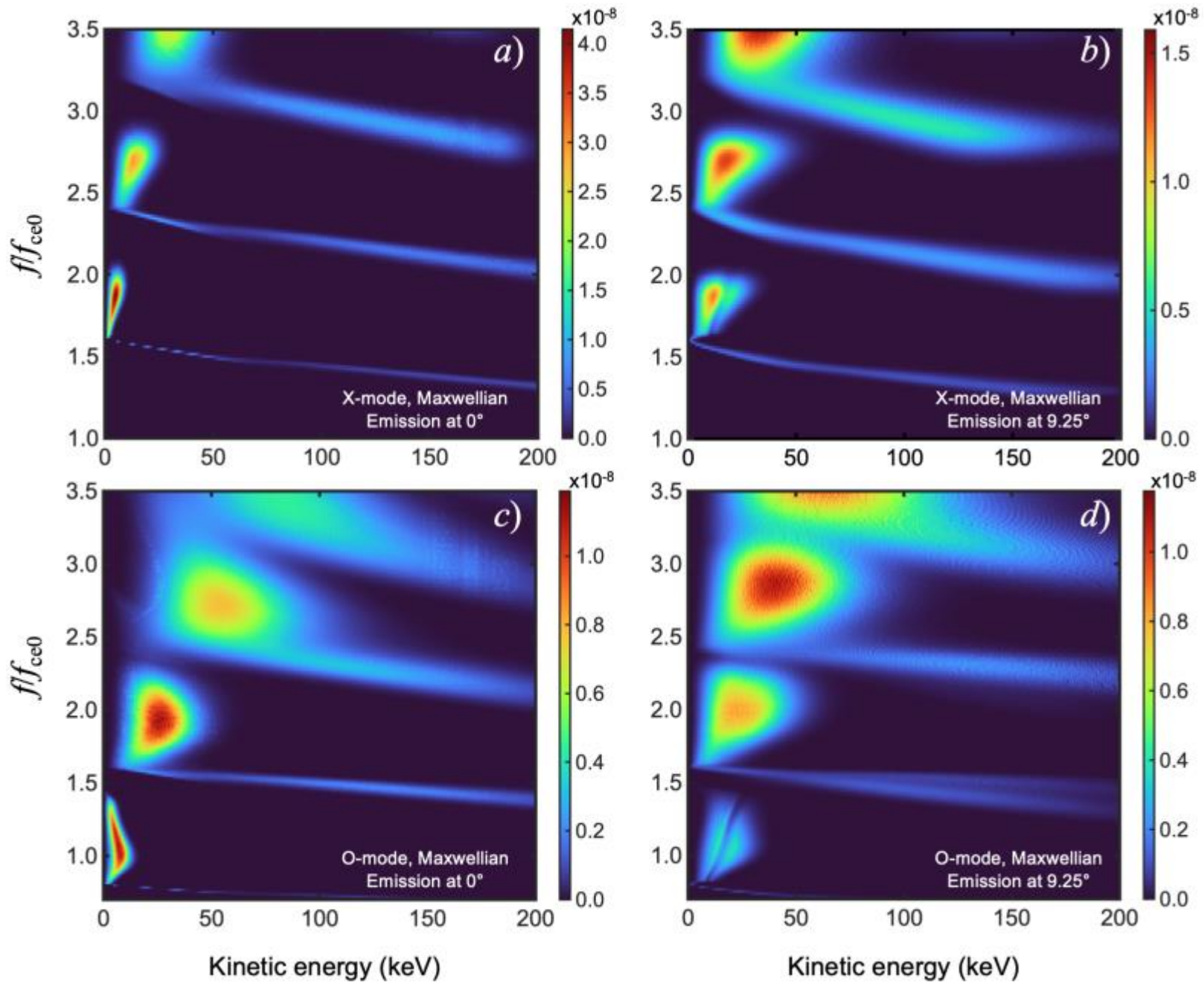

**Figure 4.** Emission flux intensity plots with Maxwellian EVDF at all radii, shown versus emission frequency normalized to the on-axis fundamental cyclotron frequency ($f_{ce0}$ = 148.4 GHz) and electron kinetic energy measured at the plasma edge, for: *a*) X-mode polarization at the ITER ECE radial view, *b*) X-mode polarization at the ITER ECE oblique view, *c*) O-mode polarization at the ITER ECE radial view, and *d*) O-mode polarization at the ITER ECE oblique view.

Figure 4 shows that for a Maxwellian EVDF, the X-mode radial view is sensitive to electron energies below ~10 keV at the second harmonic and up to ~25 keV at the third harmonic. The X-mode oblique view extends this sensitivity, reaching ~25 keV at the second harmonic and ~50 keV at the third. For O-mode, the radial view responds to electrons below ~15 keV at the first harmonic and up to ~50 keV at the second harmonic. The oblique O-mode view further extends the accessible energy range, reaching ~40 keV at the first harmonic and ~65 keV at the second.

## 2.1. Sensitivity of the Oblique ECE View to Core EVDF Distortions

To evaluate how non-thermal electrons impact the performance of the oblique ECE view, we introduce a controlled distortion into an otherwise Maxwellian electron velocity distribution function (EVDF) and compute the corresponding ECE spectra. The simulations use the same kinetic profiles and equilibrium as introduced earlier. The distorted EVDF is modeled as an isotropic, relativistic two-temperature distribution with modifications spanning $0.75 < u/u_{th} < 1.5$, as it is shown in Figure 5: this perturbation is applied only at the core radius $\rho = 0.1$ and the EVDFs at all other radii remain Maxwellian.

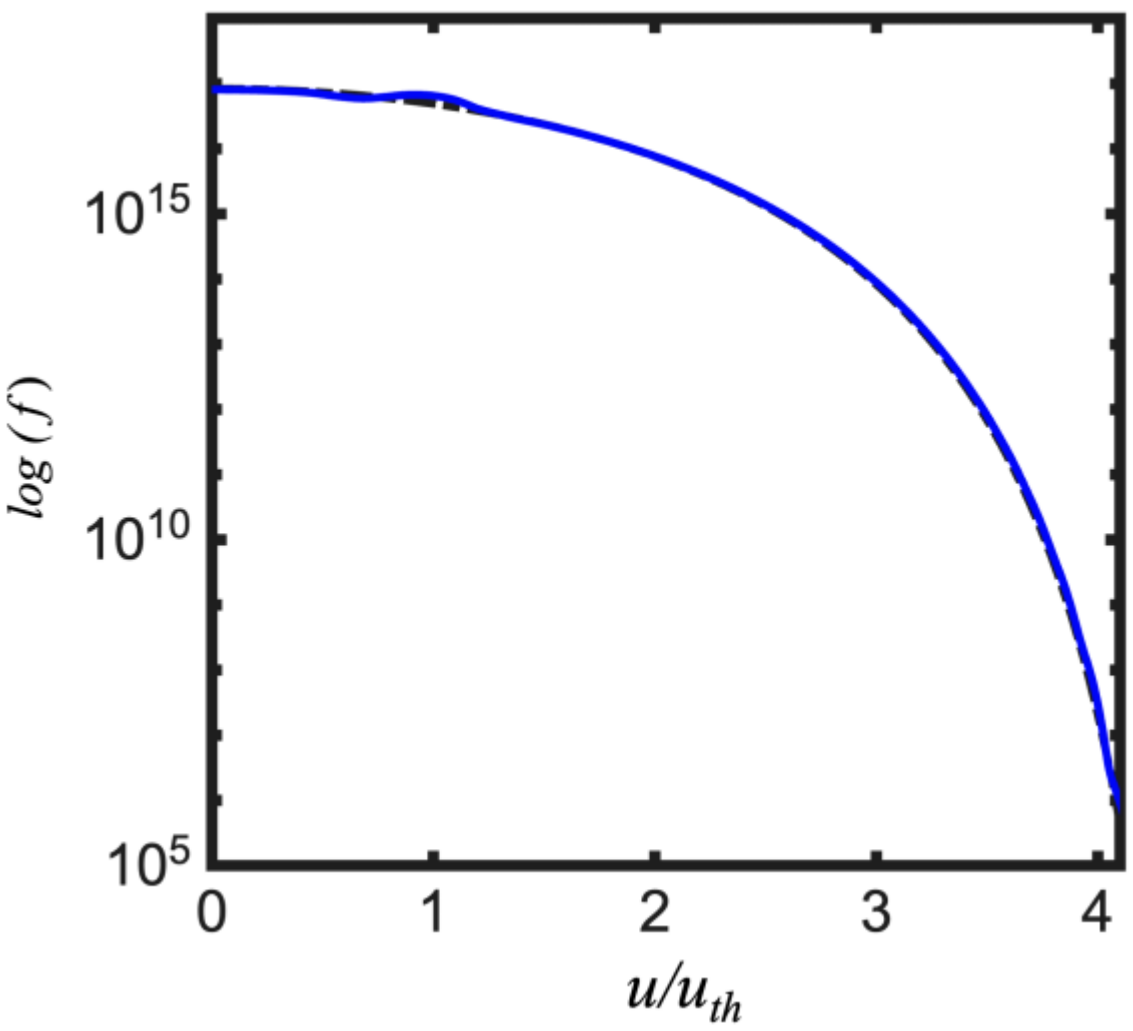

**Figure 5.** Distorted electron momentum distribution function (solid line) compared to the Maxwellian distribution (dashed line), plotted as a function of momentum normalized to the thermal momentum ($u/u_{th}$). Only the EVDF of the core region ($\rho$ = 0.1) was modified, while distributions at other radii remain Maxwellian.

Figure 6.*a*, 6.*b*, 6.*d*, and 6.*e* show the emission-flux intensity plots corresponding to the non-Maxwellian EVDF shown in Figure 5. The imposed distortion produces a clear spectral signature: a pronounced emission enhancement at the second X-mode harmonic and a symmetric double-peaked structure at the first O-mode harmonic. In the oblique view, however, non-thermal features at the third harmonic of X-mode as well as the second harmonic of O-mode are largely suppressed. This reduction can be advantageous for $T_e$-profile reconstruction, as higher harmonics become less affected by non-thermal emissions and therefore yield more stable and reliable measurements under ITER-relevant conditions.

The corresponding $T_{rad}$ spectra are shown in Figure 6.*c* and *f*. Here, simulated traces for the distorted EVDF are compared with Maxwellian spectra for both radial and oblique views. Consistent with the emission-flux intensity plots, the radial view shows strong non-thermal signatures: an asymmetric enhancement on the low-field side ($1.6 < f/f_{ce0} < 2.0$) in the second-harmonic X-mode, and symmetric broadening around the fundamental frequency in the first-harmonic O-mode. In contrast, the oblique view in both polarizations remains largely unaffected, a consequence of Doppler broadening dominating over relativistic and non-thermal effects at this geometry. At higher harmonics, non-thermal emissions become negligible, further supporting the use of oblique-view higher-harmonic ECE measurements for robust $T_e$ diagnostics.

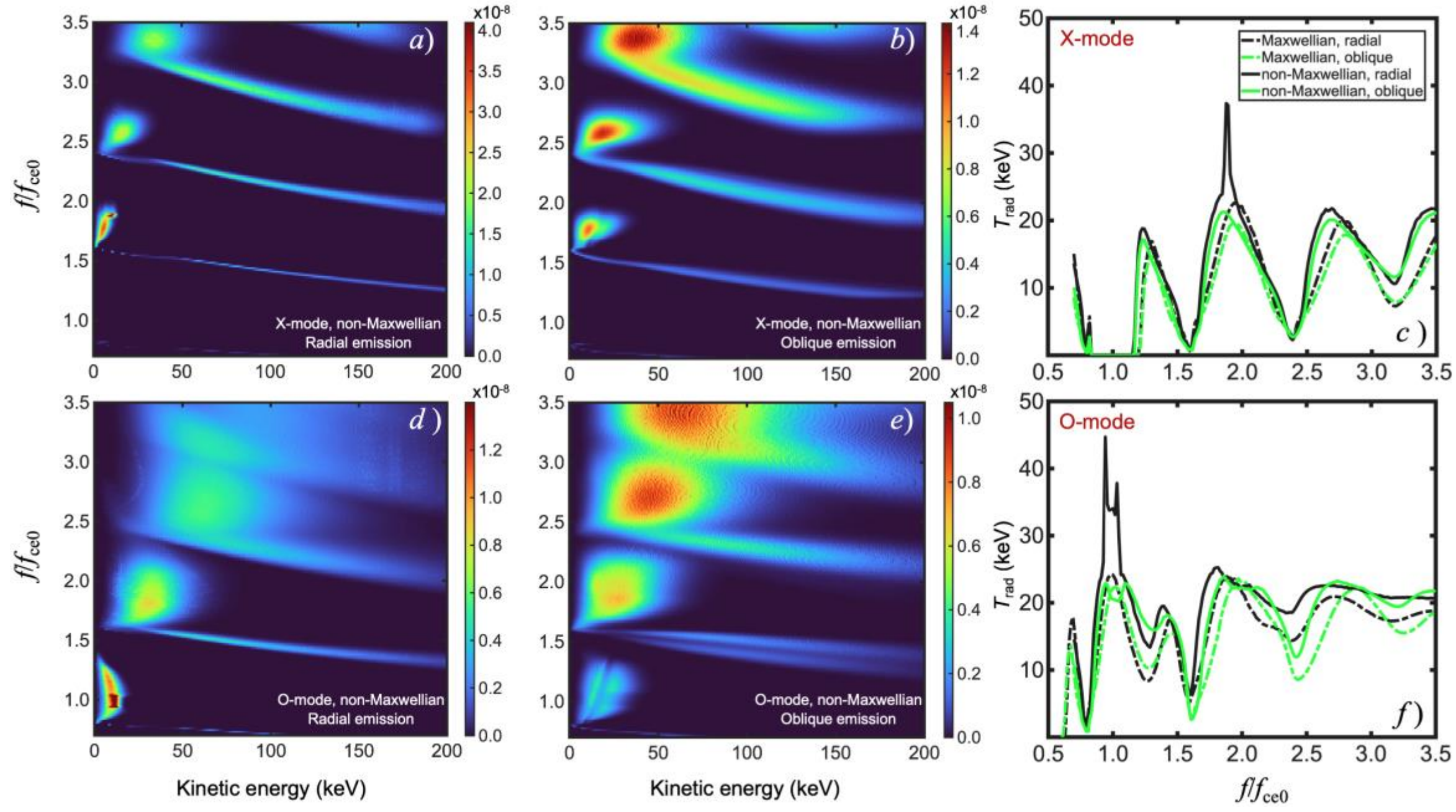

**Figure 6.** Emission flux intensity plots for the non-Maxwellian EVDFs shown in Figure 5, presented as functions of emission frequency normalized to the on-axis fundamental cyclotron frequency ($f_{ec0}$ = 148.4 GHz) and electron kinetic energy at the plasma edge, for: *a*) X-mode polarization, radial view, *b*) X-mode polarization, oblique view, *d*) O-mode polarization, radial view, *e*) O-mode polarization, oblique view. Simulated ECE spectra for the same non-Maxwellian EVDF are shown separately in panels *c*) X-mode spectra for radial (solid) and oblique (dashed) views and *f*) O-mode spectra for radial (solid) and oblique (dashed) views. In both panels *c* and *f*, spectra computed using a Maxwellian EVDF are included as references.

In the earlier analysis, a non-Maxwellian distribution was created by manually introducing a distortion into the original Maxwellian EVDF at a fixed location in momentum space. To extend this work and determine whether the distortion's location in the momentum space influences non-thermal emission, the position of the distortion was systematically scanned. As shown in Figure 7, this produced a series of two-temperature Maxwellian EVDFs, with the distortion location varied from low $u/u_{th}$ to higher values. The figure also marks the positions of the second and third EC harmonic resonance layers in momentum space. In total, 13 distorted EVDFs were generated. Although these distortions exceed the levels typically produced by heating actuators in a real plasma, they are intentionally exaggerated for simulation purposes to evaluate how sensitive the ECE spectra are to the distortion's location in momentum space.

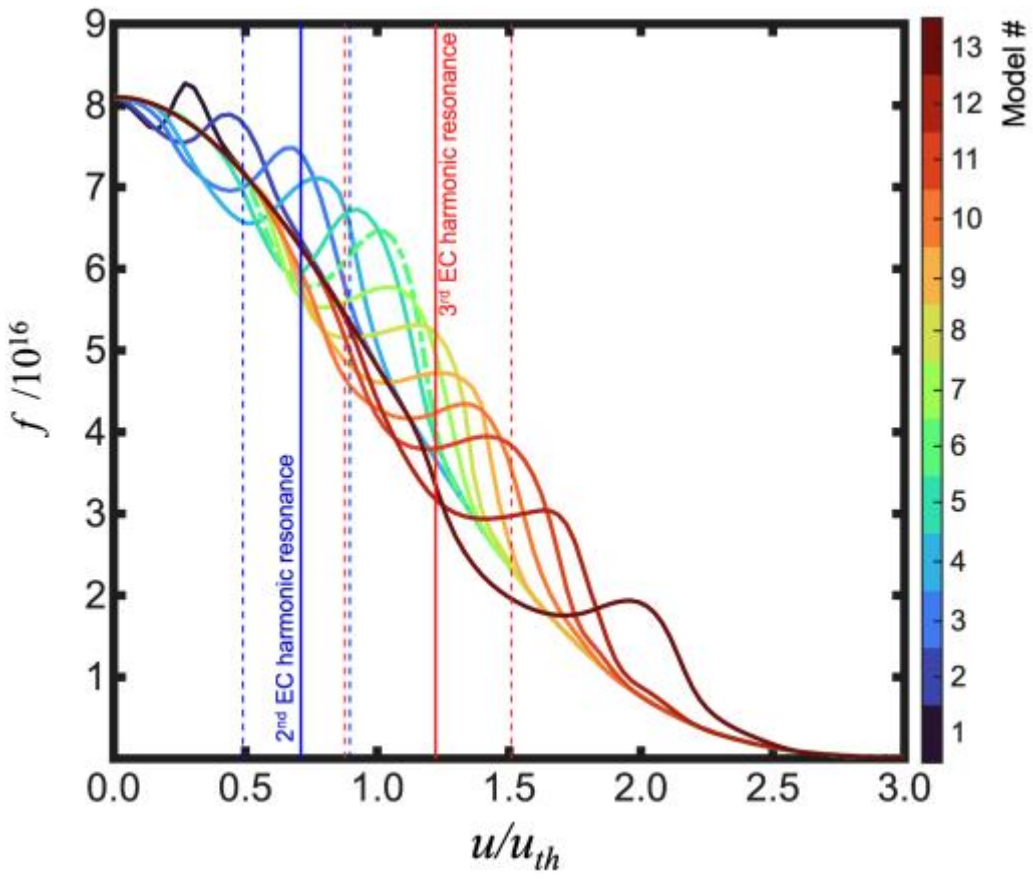

**Figure 7.** Thirteen EVDF models with localized distortions applied at different locations in normalized momentum space at the plasma core ($\rho$ = 0.1), along with the positions of the second and third EC harmonics.

Figure 8 shows the X-mode emission-flux intensity plots for models 6 and 11 from Figure 7, viewed both radially and obliquely. These two models contain distortions to the background Maxwellian EVDF located near the second and third EC harmonics, respectively. Comparing panels (*a*) with (*c*) and (*b*) with (*d*) indicates that the oblique ECE view is largely insensitive to the non-thermal features produced by these distortions, as highlighted by the dotted circles. Additionally, comparing panels (*a*) with (*b*) and (*c*) with (*d*) shows that when a distortion is placed at one EC harmonic, the higher harmonics remain unaffected. This illustrates that non-thermal modifications to the EVDF influence primarily the harmonic at which the distortion occurs, with minimal impact on adjacent harmonics.

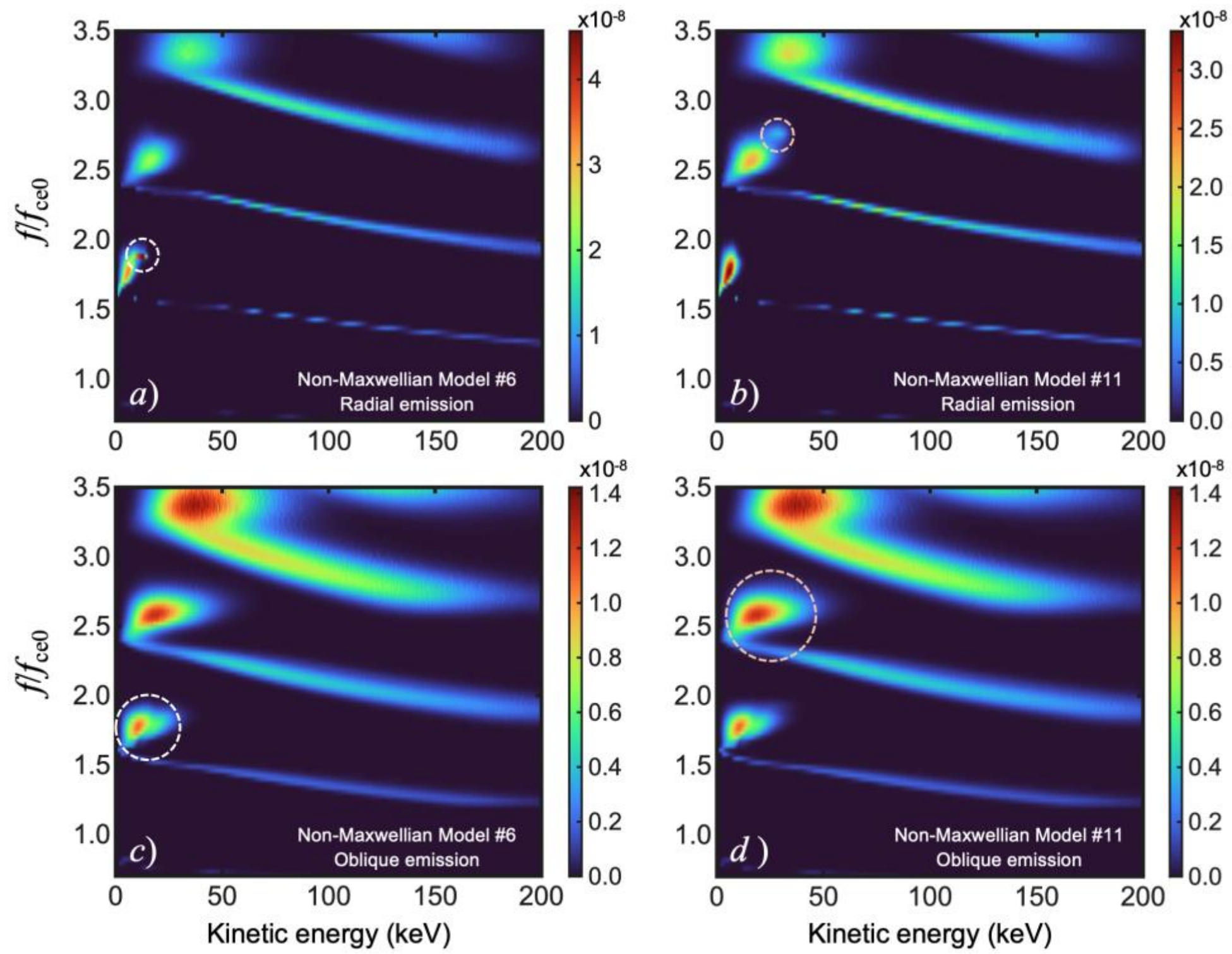

**Figure 8.** X-mode emission-flux spectra computed for the EVDF models shown in Figure 7. Panels illustrate spectra for two non-Maxwellian EVDFs model 6 and model 11, viewed from the radial view (*a* and *b*) and from the 9.25° oblique view (*c* and *d*). The locations of non-thermal emissions are highlighted by dotted circles.

Shown in Figure 9 are the simulated ECE spectra for each of these models. The distortions were applied only to the EVDF at the plasma core ($\rho$ = 0.1). As shown in Figure 9, the simulated $T_{\mathrm{rad}}$ at the core is significantly affected only when the distortion location aligns with a resonance position. For example, the $T_{\mathrm{rad}}$ spectrum in the second harmonic is affected when the distortion coincides with the second ECE harmonic resonance layer, and similarly for the third harmonic. In addition, when the distortion is aligned with a higher-order resonance, such as the third harmonic, the simulated radiation temperature exhibits a reduction relative to the Maxwellian case. This behavior is illustrated by the dark red traces, which correspond to EVDFs distorted at the third-harmonic resonance but simultaneously show a reduction in the second-harmonic spectra. These results indicate that $T_{\mathrm{ECE}}$ may be either higher or lower than $T_{\mathrm{TS}}$, depending on the location of the EVDF distortion in momentum space.

Although the distortion is localized to $\rho$ = 0.1, its spectral impact extends to neighboring radial positions. This result indicates that discrepancies can arise in the second and third X-mode harmonics if the EVDF distortion- or, more generally, the non-Maxwellian feature - occurs at only one of these resonances. Such a situation could occur, for instance, when a heating mechanism modifies the EVDF locally in momentum space without affecting other regions.

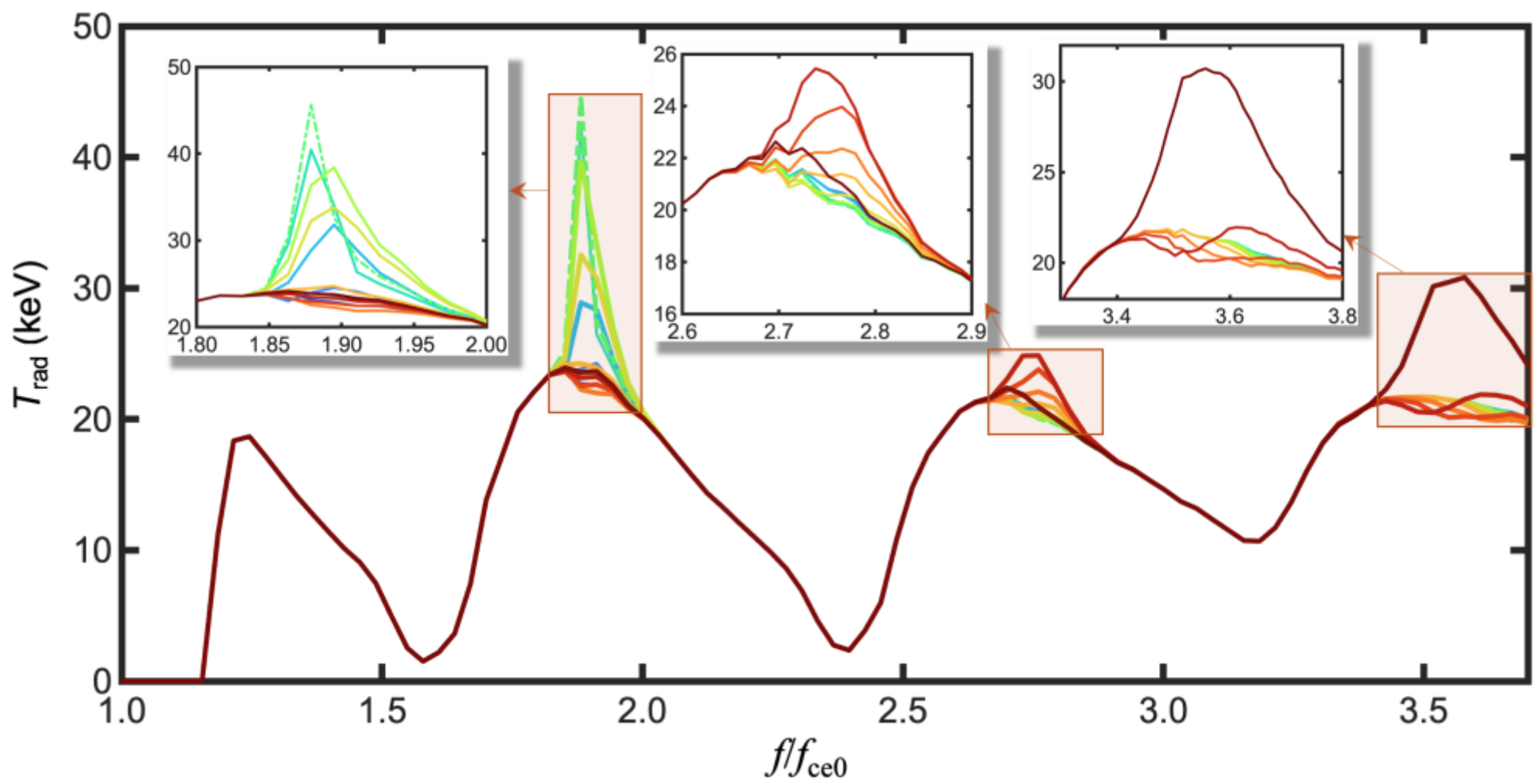

**Figure 9.** Simulated X-mode ECE spectra corresponding to the EVDFs in Figure 7, with insets highlighting the second, third, and fourth harmonics. The $T_{rad}$ corresponding to Model #6 and shown as a dashed line, produces the strongest non-thermal emission.

Figure 10 shows the intensity plot of the derivative of the 13 distribution functions from Figure 7 with respect to perpendicular velocity, each plotted against normalized momentum. The resonance layers for the second and third EC harmonics are also indicated. As noted earlier, the derivative is negative throughout most of momentum space, except in regions containing the imposed distortions in the EVDF. The plot reveals that non-thermal emission occurs only when these positive-derivative regions align with the EC harmonic resonance layers, as previously observed in Figure 9. This indicates that a positive derivative of the EVDF is necessary but not sufficient for generating non-thermal emission; it must coincide with the corresponding EC harmonic to be effective. Positive derivatives appearing at locations that do not align with EC harmonics do not produce non-thermal emission.

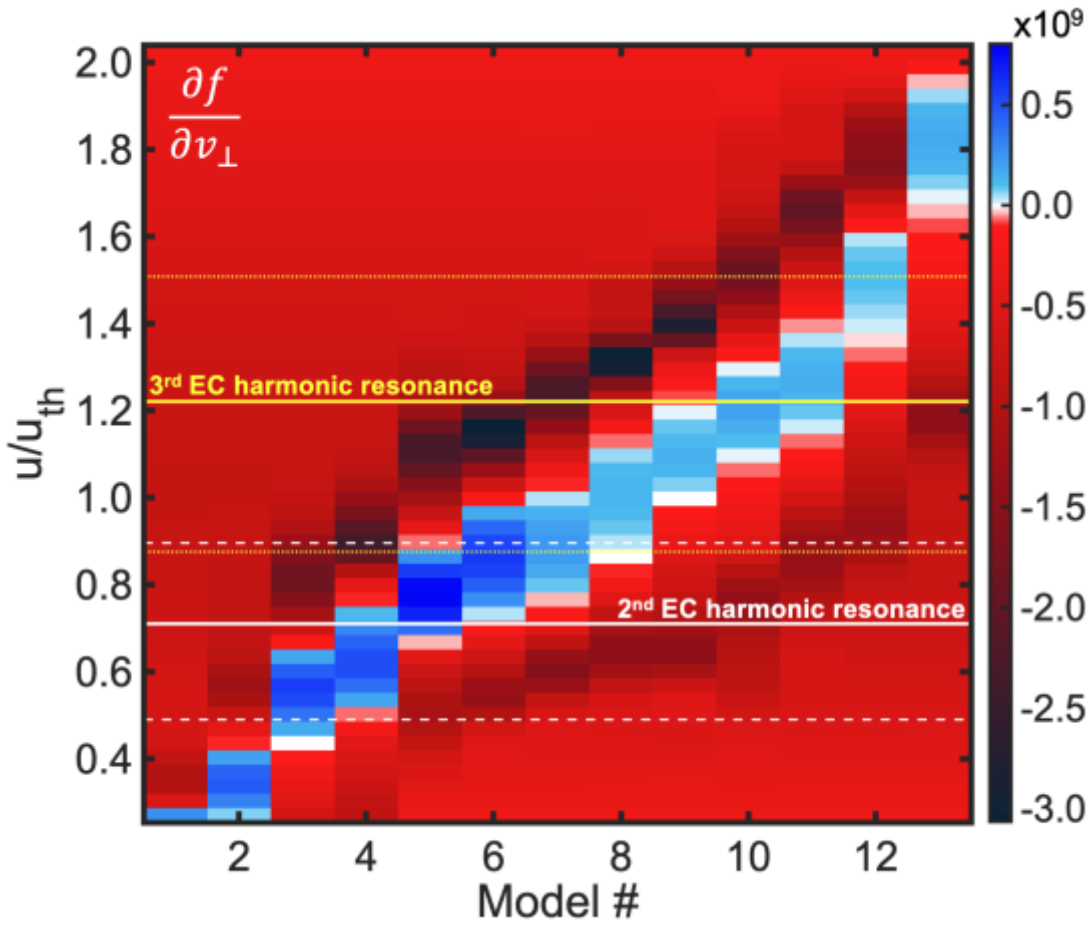

**Figure 10.** Intensity plot of the derivative of the EVDFs for each of the 13 models shown in Figure 7. The locations corresponding to the local electron temperature for the second and third EC harmonics in normalized momentum space are indicated.

The effect of the oblique viewing is then examined for the case with a single-radius EVDF at $\rho = 0.1$. To evaluate whether Doppler broadening dominates over relativistic effects, simulations were performed by scanning the viewing angle from 0° (radial view) to 10°, noting that the ITER oblique view is fixed at 9.25°. For these simulations, the EDVF Model #6 with the maximum non-thermal emission (shown with dashed traces in Figure 7 and Figure 9) was selected. Figure 11 presents the simulated ECE spectra for the second harmonic of X-mode. As the viewing angle increases, the non-thermal feature in $T_{rad}$ diminishes, indicating that Doppler broadening increasingly outweighs non-thermal effects. While Doppler broadening in the oblique geometry reduces the visibility of fine non-thermal spectral features, non-Maxwellian EVDFs can still bias inferred electron temperatures if inappropriate harmonics are used. Accurate interpretation therefore requires careful channel selection and, in some cases, synthetic diagnostic modeling.

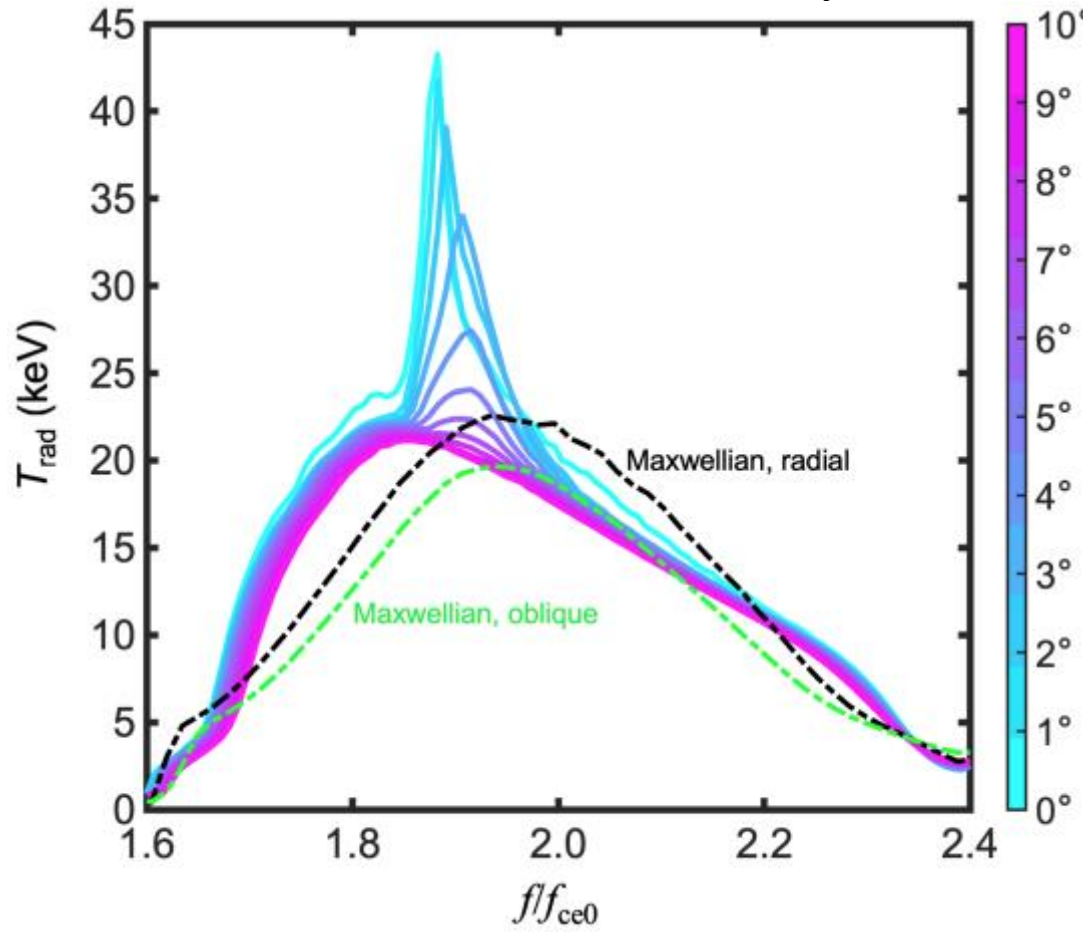

**Figure 11.** Simulated $T_{rad}$ for the second X-mode harmonic using the non-Maxwellian EVDF Model #6 shown in Figure 7. The colorbar indicates the oblique angle values used in the simulations, as labeled next to it. For reference, $T_{rad}$ from the Maxwellian EVDF at both radial and oblique views is also shown.

## 2.2. Effects of Multi-Radius Non-Maxwellian EVDFs on Oblique ECE Measurements

One possible outcome of heating mechanisms and their consequence may be the formation of EVDF distortions at multiple plasma radii. To investigate this, EVDFs at seven consecutive core positions were modified to include non-Maxwellian features. For each radius, the location of the second EC harmonic in normalized momentum space was determined. It is important to note that, because the $T_e$ varies with radius, the corresponding location of the second EC harmonic layer in normalized momentum space also changes. This approach follows directly from the findings of the previous section, which showed that a distortion must coincide with an EC resonance layer to produce non-thermal emission in the ECE spectra. The modified EVDFs at the seven radial positions are shown in Figure 12.

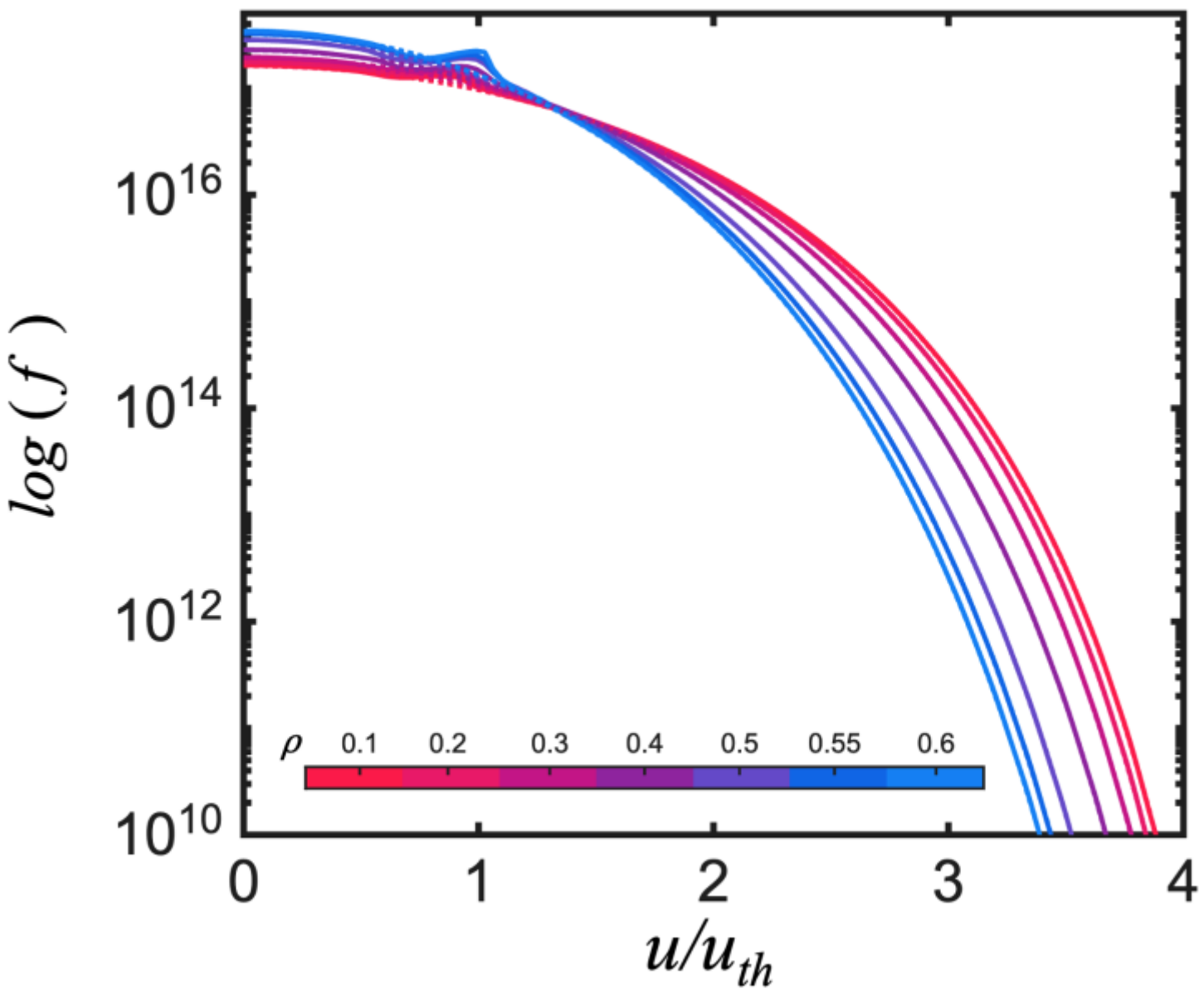

**Figure 12.** Distribution functions plotted versus normalized momentum for multiple core radii, as indicated by the inset colorbar. The corresponding Maxwellian EVDF at each radius is shown as a dotted line.

Figure 13.*a* presents GENRAY simulations of ECE spectra in the X-mode for viewing angles ranging from radial (0°) to oblique views up to 10°. Note that the ITER ECE oblique view in 9.25°. For each viewing angle, a complementary simulation was performed at the corresponding mirror angles of equal magnitude and opposite sign (*e.g.*, +1° and −1° relative to the radial view), as illustrated in Figure 13.*a*. These results show that the spectra are identical for mirror angles, confirming that the emission is symmetric with respect to the machine center for equal and opposite oblique views. Similar to the single-radius case, increasing the oblique angle causes Doppler broadening to dominate over the non-thermal effects produced by non-Maxwellian distortions in the EVDFs.

Figure 13.*b* shows the reconstructed $T_e$-profile from the second-harmonic portion of the spectrum in Figure 13.*a*, along with the corresponding profiles for a Maxwellian distribution. Although distortions were not applied beyond $\rho$ = 0.6 and no clear non-thermal emission is present in these outer regions, the reconstructed profiles yield higher $T_e$ values than the Maxwellian case. This suggests that discrepancies between ECE and Thomson scattering measurements can arise when non-Maxwellian features occur at different locations in momentum space.

Figure 13.*c* presents the reconstructed $T_e$-profile from the third-harmonic portion of the spectrum in Figure 13.*a*. The results indicate that the third-harmonic X-mode and at oblique angle is largely insensitive to non-thermal electrons arising from multiple distorted EVDFs, suggesting its potential for reliable $T_e$ profile reconstruction. Nevertheless, both Figure 13 *b* and *c* show significant offsets relative to the Maxwellian-based reconstruction. While changes in the EVDF can explain part of this ECE/TS discrepancy, it should be noted that the $T_e$-profile reconstructions here use the original magnetic equilibrium. In practice, modifications to the $T_e$-profile, such as those caused by non-thermal electrons, would require a flux-matched profile and a transport solver in an iterative process to obtain a self-consistent kinetic equilibrium. An example of such an iterative reconstruction process can be found in Ref. 27.

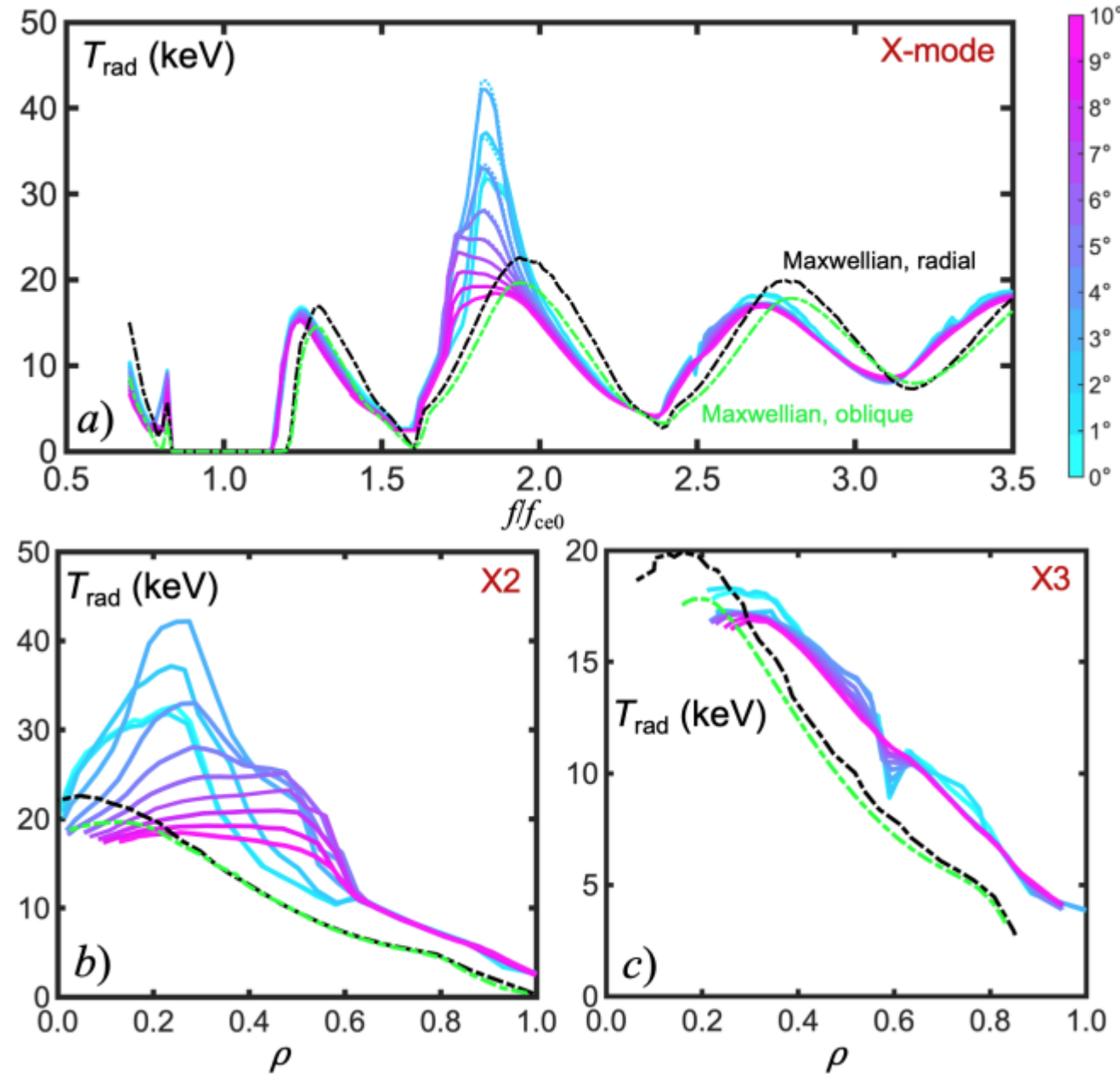

**Figure 13.** *a*) Simulated ECE spectra for X-mode emissions with oblique viewing angles scanned from 0° to 10°, incorporating the distorted EVDF shown in Figure 12. Dotted traces indicate spectra for the corresponding negative angles (*e.g.*, -1°, -2°, …). Reference spectra simulated using Maxwellian EVDFs for the ITER ECE radial (0°) and oblique (9.25°) viewing angles are also shown for comparison. *b*) Reconstructed $T_e$ profiles from the spectra in part *a* for the second X-mode harmonic range. *c*) Reconstructed $T_e$-profiles from the spectra in part *a* for the third X-mode harmonic range. In *b* and *c*, simulated $T_e$-profiles from the ITER ECE radial (0°) and oblique (9.25°) views with a Maxwellian distribution are included for reference.

Figure 14.*a* presents GENRAY simulations of ECE spectra in the O-mode for viewing angles ranging from radial (0°) to oblique views up to 10°. For each viewing angle, a complementary simulation was performed at the corresponding mirror angles of equal magnitude and opposite sign (*e.g.*, +1° and −1° relative to the radial view), as shown in Figure 14.*a*. Similar to the ECE spectra of the X-mode, these results confirm that the spectra are identical for mirror angles, indicating that emission is symmetric with respect to the machine center for equal and opposite oblique views.

As in the single-radius case, and similar to the X-mode spectra, increasing the oblique angle causes Doppler broadening to dominate over non-thermal effects from non-Maxwellian distortions in the EVDFs. However, unlike the X-mode spectra, non-thermal emissions in the O-mode appear symmetrically around the fundamental cyclotron frequency in the first harmonic. Figure 14.*b* shows the reconstructed $T_e$ profile from the first-harmonic portion of the spectrum in Figure 14.*a*, along with the corresponding Maxwellian-based profiles. Similar to the X-mode case, where $T_e$-profiles were reconstructed from the second X-mode harmonic, distortions limited to $\rho \leq 0.6$ and the absence of clear non-thermal emission in the outer radii still result in reconstructed profiles with higher $T_e$ than the Maxwellian case. Unlike the X-mode, however, non-thermal emission appears on both the low-field and high-field sides.

Figure 14.*c* shows the reconstructed $T_e$-profile from the second-harmonic portion of the O-mode spectrum in Figure 14.*a*. The modeled second-harmonic O-mode remains sufficiently optically thick to support ECE-based Te inference, as reflected in the absorption profiles computed by GENRAY. The results suggest that the second-harmonic O-mode is largely unaffected by non-thermal electrons from multiple distorted EVDFs. Notably, there is strong agreement between the core $T_e$-profiles reconstructed from Maxwellian and non-Maxwellian EVDFs at the oblique view, indicating that this harmonic may be suitable for core $T_e$ measurements. Nonetheless, Figure 14 *b* and *c* both show profile shifts, particularly beyond mid-radius relative to the Maxwellian reconstruction, which could be addressed through kinetic equilibrium reconstruction, which would then be used for accurate ECE mapping.

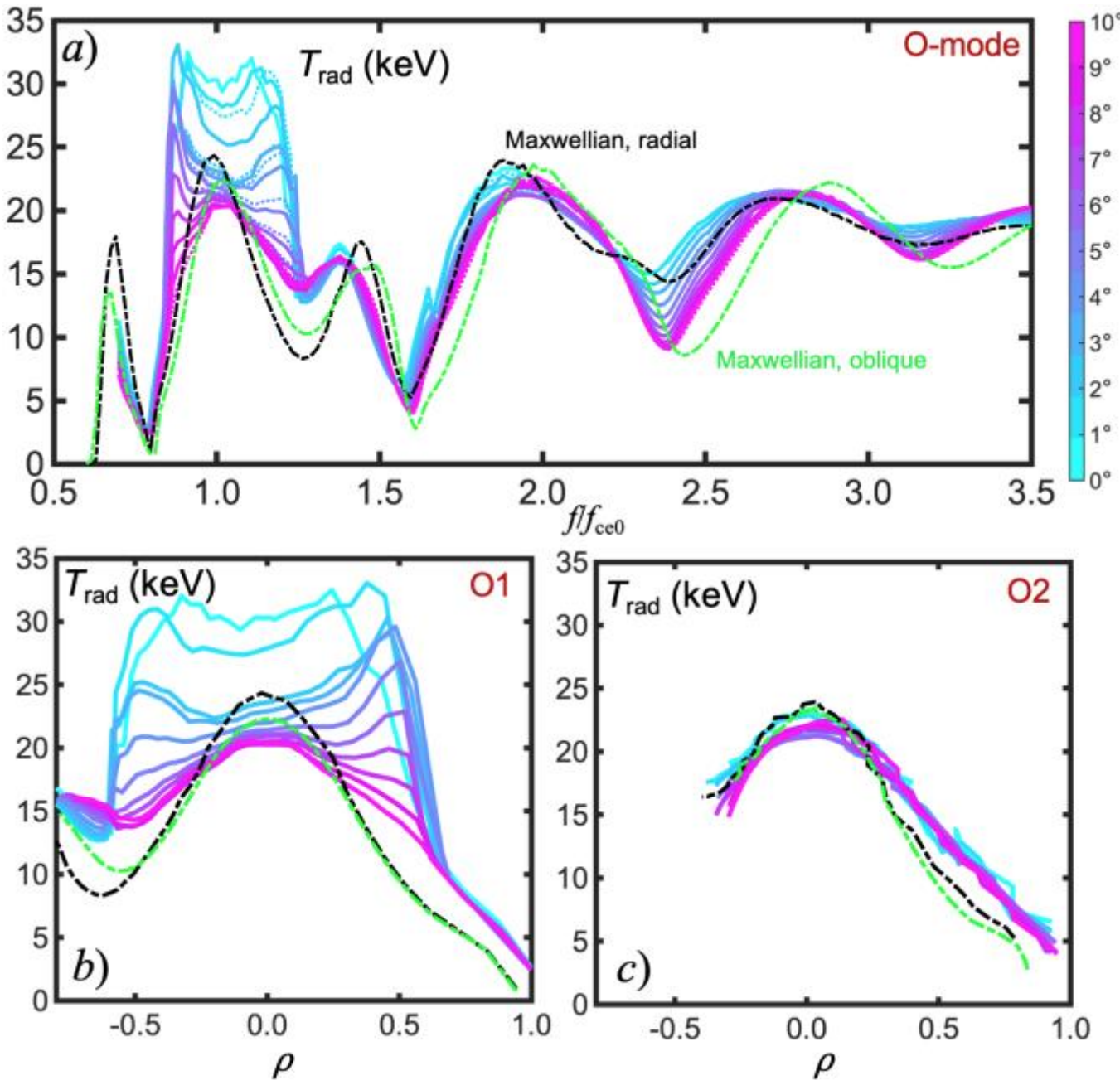

**Figure 14.** *a*) Simulated ECE spectra for O-mode emissions with oblique viewing angles scanned from 0° to 10°, incorporating the distorted EVDF shown in Figure 12. Dotted traces indicate spectra for the corresponding negative angles (*e.g.*, -1°, -2°, …). Reference spectra simulated using Maxwellian EVDFs for the ITER ECE radial (0°) and oblique (9.25°) viewing angles are also shown for comparison. *b*) Reconstructed $T_e$ profiles from the spectra in part *a* for the first O-mode harmonic range. *c*) Reconstructed $T_e$-profiles from the spectra in part *a* for the second O-mode harmonic range. In *b* and *c*, simulated $T_e$-profiles from the ITER ECE radial (0°) and oblique (9.25°) views with a Maxwellian distribution are included for reference.

## 3. Conclusions

This study used GENRAY 3D ray-tracing simulations to evaluate the performance of the ITER ECE diagnostic system's 9.25° oblique view under ITER-relevant plasma conditions. Simulations were carried out for both Maxwellian and non-Maxwellian EVDFs, the latter incorporating localized two-temperature distortions at the plasma core. Results show that the second-harmonic X-mode in the oblique view remains sensitive to electron kinetic energies up to ~50 keV, and that higher harmonics in both X- and O-mode retain robust spectral features even in the presence of EVDF distortions. This enables reliable temperature profile reconstruction across a wide range of conditions. Angle scans from 0° to 10° revealed that Doppler broadening increasingly dominates over relativistic effects at larger oblique angles, which reduces sensitivity to fine non-Maxwellian features but enhances the stability of temperature profile measurements. These findings confirm that the implemented 9.25° oblique view meets its design objectives, providing both sensitivity to non-thermal electrons and robustness in electron temperature diagnostics for reactor-grade plasmas. These findings provide a framework for interpreting ECE measurements in the presence of non-Maxwellian EVDFs and help explain how ECE-TS discrepancies may arise or be mitigated through appropriate choice of viewing geometry and harmonic.

An important operational implication of this study is that synthetic, scenario-specific ECE simulations should be carried out prior to ITER plasma operation to guide the selection of polarization and harmonic channels for temperature measurements. This does not require prior knowledge of the actual electron temperature profile during a given discharge. Rather, it leverages reference ITER operating scenarios, such as those developed within the IMAS framework, to anticipate how the diagnostic system is expected to respond under reactor-relevant conditions. By comparing these precomputed spectral responses with measurements obtained during operation, it becomes possible to identify which viewing geometry, harmonic, and polarization provide the most reliable basis for $T_e$ inference, particularly in the presence of non-thermal electrons. This consideration has direct implications for the current ITER ECE system design, which includes a high-frequency radiometer, a low-frequency radiometer, and two Fourier Transform Spectrometers (FTSs). With four possible measurement configurations (radial X-mode, radial O-mode, oblique X-mode, and oblique O-mode), only two are presently accessible at any given time, leaving the remaining configurations unused.

The proposed switchyard in the ITER ECE design aims to select between these configurations, but the results of this study strongly support equipping all four channels with FTS capability. This would allow real-time access to the full ECE spectrum for all polarizations as well as the polarizations harmonics, enabling data-driven optimization of radiometer settings and ensuring that critical non-Maxwellian features are not overlooked. Significant design attention should therefore be given to expanding the switchyard concept to fully utilize all four ECE measurement paths.

## 4. Appendix I

Using cold plasma dispersion, the ECE channel frequencies can be calculated. The location of each ECE channel is mapped to its radial position in the tokamak using Equation (A.1).

$$\omega_n = n\frac{e}{\gamma m_e}|\mathbf{B}|\frac{1}{1-v_{\parallel}cos\theta/c}. \tag{A.1}$$

In Equation (A.1), $\gamma = (1 - v^2/c^2)^{-1/2}$ is the Lorentz factor, $c$ is the speed of light, **B** represents the total magnetic field vector, composed of radial, azimuthal, and vertical components. These components become available once the magnetic equilibrium is reconstructed. As an approximation, the field magnitude can be expressed as $|\mathbf{B}| \approx B_0R_0/R$, where $B_0$ is the magnetic field on axis, $R_0$ is the major radius, and $R$ is the distance from the magnetic axis. This expression provides an excellent estimate of the field strength obtained from the reconstructed magnetic equilibrium.

Also, in Equation (A.1), $n$ denotes the harmonic number of the emission, and $\theta$ is the wave propagation angle. The oblique ECE view angle is defined as $(90° - \theta)$. The denominator in Eq. (A.1) captures the Doppler effects, which leads to spectral broadening of the ECE signal. This term becomes zero for a purely radial view ($\theta = 90°$). In contrast, the numerator term, $(1 - v^2/c^2)^{1/2}$, accounts for relativistic effects, resulting in both broadening and a down-shift of the ECE spectrum. However, in the design and positioning of ECE channels, the corresponding radial location is typically determined using the non-relativistic cold plasma approximation: $ne|B|/m_e$. The radii derived from this relation are commonly referred to as $R_{cp}$, representing the ECE radial position based on cold plasma approximation. Post-processing of the ECE profile includes identifying the relativistically corrected emission layer, $R_{rel}$, as well as determining the width of the emission layer, as described in the following paragraphs.

Following the ECE review study summarized by Bornatici,[28,29] the relativistic broadening and frequency shift of the emission profile are calculated here to quantify the deviation from $R_{cp}$. The procedure involves identifying the peak of the relativistically broadened and shifted emission profile. The density and temperature profiles shown in Figure 2 are used for this analysis. At each radial location, the emission profile is calculated as follows:

$$G(s) = T_e(s) \cdot \alpha(s) \cdot e^{-\tau(s)}. \tag{A.2}$$

In Eq.A.2, $T_e$ is the electron temperature, $\alpha$ is the absorption coefficient, $\tau = \int \alpha(\omega,s)\, ds$, is the optical depth, $\omega$ is the angular frequency of the wave and $s$ is distance along the ray trajectory originating from the point of interest within plasm extended to the antenna at the plasma edge. The emission and absorption process in plasma is described by the following equation:

$$N_r^2 \hat{s}.\nabla\left(\frac{I}{N_r^2}\right) = j - \alpha I. \tag{A.3}$$

Here $\hat{s}$ is the unit vector along the ray trajectory ($s$), $N_r$ is the ray refractive index, $j$ is emissivity and $I$ is the radiance. The solution of the Eq. A.3 can be written as:

$$I(\omega, s) = I_0 e^{-\tau(s)} + \int \frac{j(\omega,s)}{N_r(\omega,s)^2} e^{-\tau(s)} ds, \tag{A.4}$$

where $I_0$ is the incident radiation intensity, which is negligible for ECE measurements. With a Maxwellian velocity distribution and a plasma in thermodynamic equilibrium and blackbody approximation in the Raleigh-Jeans limit where

$$G(s) = \frac{j(\omega,s)}{N_r(\omega,s)^2} = \frac{\omega^2}{8\pi^3 c^2} T_e(s) \cdot \alpha(\omega, s), \tag{A.5}$$

it can be shown[30] that the radiance can be written as:

$$I(\omega, s) = \frac{\omega^2}{8\pi^3 c^2} kT_e = \frac{\omega^2}{8\pi^3 c^2} k \int G(s) ds. \tag{A.6}$$

For $n^{th}$ harmonic of extraordinary mode of propagation (X-mode), the absorption coefficient is defined as

$$\alpha_n^X = A_n \frac{n^{2n-1}}{2^n n!} \left(\frac{\omega_{pe}}{\omega_{ce}}\right)^2 \left(\frac{T_e}{m_e c^2}\right)^{n-2} \frac{\omega_{ce}}{c} \left[-F''_{n+\frac{3}{2}}(Z_n)\right] \tag{A.7}$$

where,

$$A_n = N_\perp^{2n-3} \left|1 + \frac{\left(\frac{\omega_{pe}}{\omega_{ce}}\right)^2}{n\left[n^2 - 1 - \left(\frac{\omega_{pe}}{\omega_{ce}}\right)^2\right]}\right|^2, \ F''_q = -\frac{\pi}{\Gamma(q)} |Z_n|^{q-1} e^{-|Z_n|}, \ Z_n = \left(\frac{c}{v_{th}}\right)^2 \left(\frac{n\omega_c}{\omega} - 1\right)$$

Here, $\omega_p$ and $\omega_{ce}$ are plasma and cyclotron frequencies, $N_\perp$ is the perpendicular component of the refractive index and $v_{th}$ is the electron thermal velocity. Note that the emissivity at each radius depends on the local magnetic field, temperature, and density. Therefore, it is important to emphasize that the calculation of relativistic corrections is a post-discharge process. The actual emission layer, $R_{rel}$, is identified as the location where the emissivity function, $G$, reaches its maximum. To determine the width of the emission layer, the integral of $G$ is calculated, and the bounds are defined by the positions where the cumulative emissivity reaches 5% and 95% of the total, denoted R05 and R95, respectively. When representing this layer as a data point with error bars in plots, the left half-error bar is defined as the distance between $R_{rel}$ and R05, and the right half-error bar as the distance between $R_{rel}$ and R95. An example of the absorption, optical depth and the emissivity functions are shown in Figure 15.

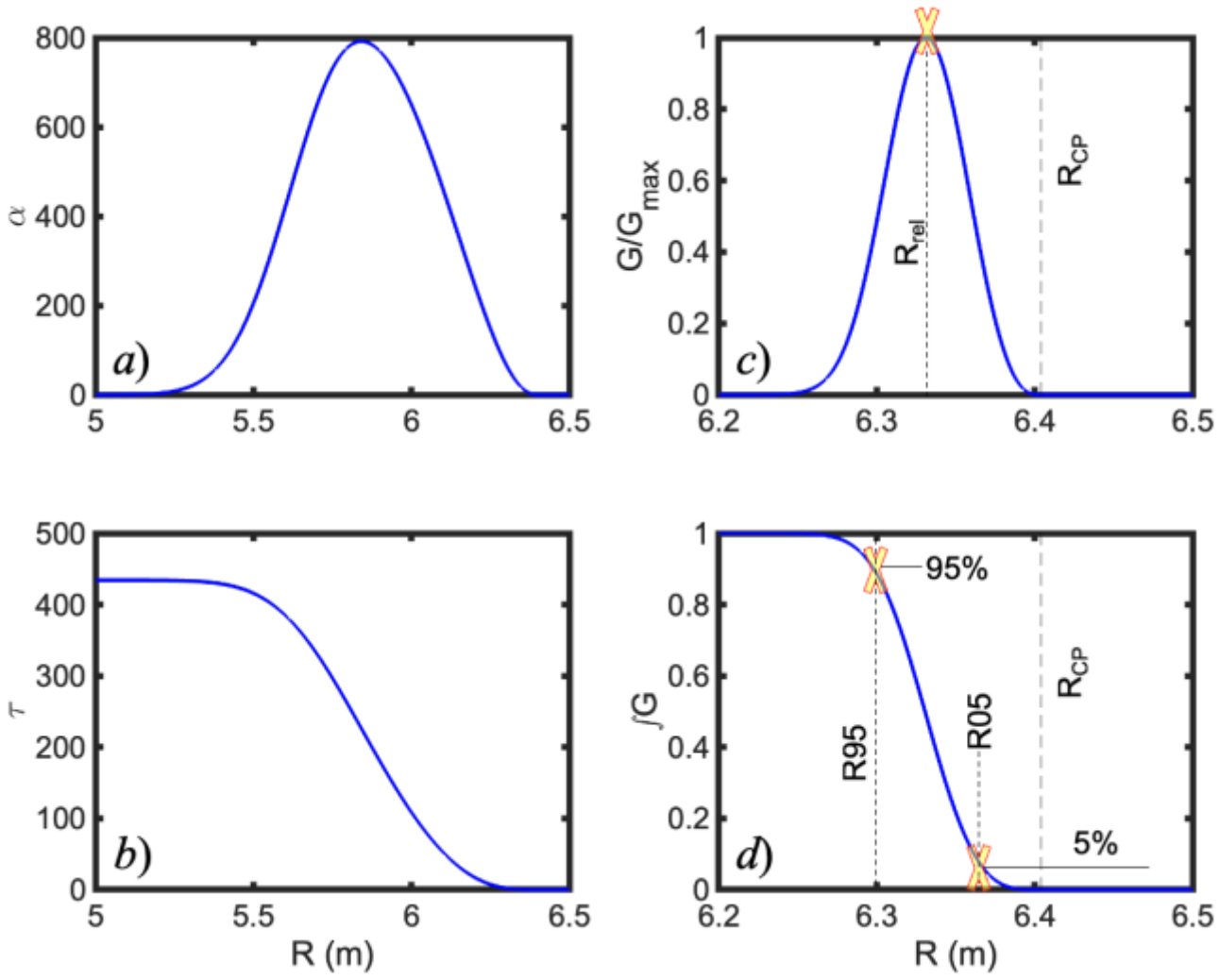

**Figure 15.** Profiles of *a*) absorption coefficient, *b*) optical depth, and *c*) emissivity function for a single 2nd harmonic X-mode frequency of 290 GHz in ITER in the case shown in figure 1. In panel *d* the emission width is determined as the distance between 5% and 95% emission levels, and the center of the emission is demarked by the 50% radius.

## Acknowledgments

The This work is supported PPPL subcontract S013464-C via U.S. DOE Contract No. DE-AC02-09CH11466 with Princeton University.

## Disclaimers

The views and opinions expressed herein do not necessarily reflect those of the ITER organization. Data available on request from authors.

## ORCID IDs

S. Houshmandyar, https://orcid.org/0000-0002-4738-3569
W. L. Rowan, https://orcid.org/0000-0001-8793-919X